\documentclass[12pt]{article}
\usepackage[utf8]{inputenc}
\usepackage{geometry}
\usepackage{amsmath}
\usepackage{amssymb}
\usepackage{upgreek}
\usepackage{algorithm2e}
\usepackage{setspace}
\usepackage{amsmath}
\usepackage{makecell}
\usepackage{multirow}
\usepackage{adjustbox}
\usepackage{url}
\newcommand{\mb}{\mathbf}
\newcommand{\bs}{\boldsymbol}

\DeclareMathOperator*{\argmin}{arg\,min}

\DeclareMathOperator{\tr}{tr}

\DeclareMathOperator{\EX}{\mathbb{E}}% expected value

\usepackage[compact]{titlesec}
\titlespacing{\section}{0pt}{2ex}{1ex}
\titlespacing{\subsection}{0pt}{1ex}{0ex}
\titlespacing{\subsubsection}{0pt}{0.5ex}{0ex}

\usepackage{natbib}
\usepackage{graphicx}

\usepackage{xr}
\makeatletter
\newcommand*{\addFileDependency}[1]{
  \typeout{(#1)}
  \@addtofilelist{#1}
  \IfFileExists{#1}{}{\typeout{No file #1.}}
}
\makeatother

\newcommand*{\myexternaldocument}[1]{
    \externaldocument{#1}
    \addFileDependency{#1.tex}
    \addFileDependency{#1.aux}
}
\myexternaldocument{supplementary}

\begin{document}

\thispagestyle{empty} \baselineskip=28pt \vskip 5mm
\begin{center} {\Huge{\bf Calibration of Spatio-Temporal Forecasts from Citizen Science Urban Air Pollution Data with Sparse Recurrent Neural Networks}}
\end{center}

\baselineskip=12pt \vskip 10mm

\begin{center}\large
Matthew Bonas\textsuperscript{1} and Stefano Castruccio\footnote[1]{
\baselineskip=11pt Department of Applied and Computational Mathematics and Statistics,
University of Notre Dame, Notre Dame, IN 46556, USA.}
\end{center}

\baselineskip=17pt \vskip 10mm \centerline{\today} \vskip 15mm

\begin{center}
{\large{\bf Abstract}}
\end{center}
With their continued increase in coverage and quality, data collected from personal air quality monitors has become an increasingly valuable tool to complement existing public health monitoring systems over urban areas.  However, the potential of using such `citizen science data' for automatic early warning systems is hampered by the lack of models able to capture the high resolution, nonlinear spatio-temporal features stemming from local emission sources such as traffic, residential heating and commercial activities. In this work, we propose a machine learning approach to forecast high frequency spatial fields which has two distinctive advantages from standard neural network methods in time: 1) sparsity of the neural network via a spike-and-slab prior, and 2) a small parametric space. The introduction of stochastic neural networks generates additional uncertainty, and in this work we propose a fast approach for ensure that the forecast is correctly assessed (calibration), both marginally and spatially. We focus on assessing exposure to urban air pollution in San Francisco, and our results suggest an improvement of over 58\% in the mean squared error over standard time series approach with a calibrated forecast for up to 5 days. 

\newpage

\section{Introduction}\label{sec:intro}

According to the World Health Organization, exposure of the population to ambient air pollution is the cause of approximately 4.2 million premature deaths globally each year, of which approximately 60,000 occur in the United States (US,  \citet{zha18}). While some fatalities can be attributed to extreme events such as wildfires, most deaths in urban environments are caused by a systematic exposure to degraded air quality from local sources such as traffic, residential heating and commercial activities \citep{Good19}. Therefore, it is of high scientific interest to develop methods to understand and predict these local spatio-temporal patterns, in order to improve early warning systems and minimize population exposure \citep{Mont05,carl20}. 

From an applied perspective, even in developed countries such as the US the high quality monitoring network maintained by the Environmental Protection Agency (EPA) is not sufficiently dense in space to capture the urban variability of air pollution. For example, in San Francisco, which will be the focus of our applications in this paper, only one EPA site is currently available. Alternative satellite data sources such as the Moderate Resolution Imaging Spectroradiometer (MODIS) are very sparse in space and time, and such optical measures are not directly relatable to ground observations \citep{lev10}. Proxy methods such as land use regression, which links air pollution to other urban covariates easier to access such as traffic, altitude, temperature or other weather variables, have a varying degree of accuracy \citep{bri00}. Numerical chemical transport models such as the Weather and Research Forecasting coupled with Chemistry (WRF-Chem, \cite{gre05}) rely instead on emission inventories at maximum resolution of a few kilometers in the US and coarser in other world regions, a scale vastly insufficient to characterize the scales of variability of urban emissions. This work aims at leveraging a more recent and substantially less scientifically investigated means to monitor urban air pollution: citizen science data. As of 2022, PurpleAir is one of the most popular air quality sensor networks, which since the commercialization of the first sensors in 2015 has seen an exponential increase in number of privately owned monitoring stations in cities in the US and beyond. This work demonstrates how this new source of data can be used to produce urban maps of population exposure which can not be obtained using only EPA data. 

Our current understanding of air pollution in urban environments is that of a nonlinear process dictated by complex interactions between meteorology (wind speed, precipitation, boundary layer stability), chemical processes (primary sources and secondary formation of PM$_{2.5}$) and the built environment (street canyons and building wakes). As such, from a methodological point of view, a forecast and quantification of the spatio-temporal uncertainty of high-resolution, high-frequency air pollution at urban/local scale ($<$5km) requires accounting for the turbulent effect of local processes which cannot be effectively captured with standard dynamical models such as AutoRegressive Moving Average (ARMA), or spatial versions thereof \citep{huang2021}. More flexible approaches to characterize nonlinear dynamics have been developed in the field of machine learning, with the use of neural networks (NNs). While these \textit{recurrent NNs} (RNNs) are considerably more flexible in characterizing multivariate (possibly spatial) data resolved in time, the computational time for inference is substantial, and approaches such as explicit computation of gradients (\textit{backpropagation}) cannot be directly applied in a dynamical setting as they would lead to numerical instabilities (vanishing or exploding gradient). Additionally, a robust forecasting methodology requires an assessment and possibly an adjustment of the uncertainty to ensure that it is truly reflective of our degree of confidence in the prediction, e.g., a 95\% prediction interval must cover the true future value 95\% of the times. This \textit{calibration} of the forecast \citep{gneit07} needs to be performed both at each sampling site (marginally) and simultaneously at multiple locations (jointly) to allow a correct uncertainty quantification of predicted spatial averages or differences. Methods for assessing uncertainty in NNs (or deep versions thereof) such as dropout or dilution \citep{hin12} predicated on random deletion of nodes to assess the robustness of the network, or Bayesian approaches \citep{gra11,blu15} imply an additional computational overhead. This computational burden may be acceptable in a static problem, but unfeasible in a dynamic setting with high-frequency data where decisions need to be made in real-time.

To reduce the computational burden for both forecasting and uncertainty quantification, a modification of the RNN predicated on the use of stochastic networks is proposed. This \textit{Echo-State Network} (ESN, \cite{jae01}) still retains some fundamental properties of the RNN such as the ability to approximate an arbitrarily complex function \citep{har20,gon21}, and can be regarded as a hierarchical Bayesian model with a spike and slab prior from a statistical perspective. The advantage of a hierarchical sparse approach compared to traditional RNNs is two-fold: 1) sparse matrices allows for fast sparse linear algebra inference and 2) the parametric space is considerably smaller than all the entries of the weight matrices in standard RNNs. While ESNs have been previously used for forecasting air pollution \citep{xu19}, as well as hospitalizations from degraded air quality \citep{Arau20}, these works did not focus on quantifying the prediction uncertainty and were used on a single sampling site. In a series of recent works \citep{mcd17,mcd18,mcd19}, ESNs were used to forecast nonlinear spatio-temporal fields and the authors proposed the use of an ensemble of forecasts to quantify the uncertainty, but their approach was focused on short-term forecasting without calibration, and did not account for a natural expansion of the uncertainty through time expected in long-range forecasting. Besides the marginal forecasts, in many applications such as monitoring urban air pollution, it is of interest to characterize the spatial dependence as one might want to account for uncertainty of (local) spatial averages or differences. In cases of large spatial fields, a full analysis could prohibitively increase the computational burden of the ESN. In this case, the use of dimension reduction methods for spatial data in conjunction with ESNs was proposed in a series of recent works \citep{mcd18, mcd19,huang2021}. When the number of sampling sites is however small and spatial interpolation is of interest (as in our application with monitoring stations), an explicit characterization of the spatial dependence is necessary. 

In order to generate fast and calibrated forecasts of urban air pollution, we formulate an approach predicated on \textit{post hoc} calibration of ESN for each location individually, characterization of the long-lead uncertainty with monotonic splines as well as explicit spatial modeling with a non-stationary correlation function obtained through convolution at fixed knots. This approach allows to assess and quantify the uncertainty of air pollution forecasts up to 5 days ahead across the urban area covered by the citizen science sensors, and represent a template to provide detailed, high-resolution information about air quality and exposure, complementing large-scale, national or international efforts to monitor air pollution such as AirNow or the World Air Quality Index. 

The work proceeds as follows. Section \ref{sec:data} introduces the air pollution as well as the population data for San Francisco. Section \ref{sec:method} presents the model, along with the inferential process, the forecasting and the calibration method. Section \ref{sec:sim} presents a collection of simulation studies to assess the performance of the model in terms of point forecast and uncertainty quantification against standard time series methods. Section \ref{sec:appl} presents the results of the model in terms of air pollution forecasting and population exposure. Section \ref{sec:concl} concludes with a discussion. The code for this work is available at the following GitHub repository: github.com/MBonasND/2021ESN. 

\section{Data}\label{sec:data}

We consider concentrations of particulate matter smaller than 2.5 $\upmu$m in aerodynamic diameter (PM$_{2.5}$) from PurpleAir, one of the widest citizen science networks measuring real-time air quality conditions. The PurpleAir network uses PA-II sensors which, as of 2021, cost 249USD compared to Environmental Protection Agency (EPA) sensors, whose cost is in the range of thousands of USD \citep{ardon20}, for monitoring changes in the PM$_{2.5}$ concentrations. The PurpleAir sensors use PMSX003 laser counters to measure particulate concentrations in real time, while EPA sensors collect PM$_{2.5}$ through a filter which is weighted to determine the particulate mass concentration, a system considerably more sophisticated and expensive. The instrument precision, expressed as maximum percentage difference between two PA-II sensors inside each device has been estimated as $\pm10$\% on concentrations less than 500 $\upmu g m^{-3}$ (personal communication with the manufacturer). PA-II sensors are considered the standard for citizen science outdoor air pollution devices, and an independent study from the South Coast Air Quality Management District found overall very strong agreement ($R^2 \sim 0.93-0.97$) between these and Federal Equivalent Methods \citep{aqm21}. 

\begin{figure}[!tb]
\centering
\includegraphics[width = 11cm]{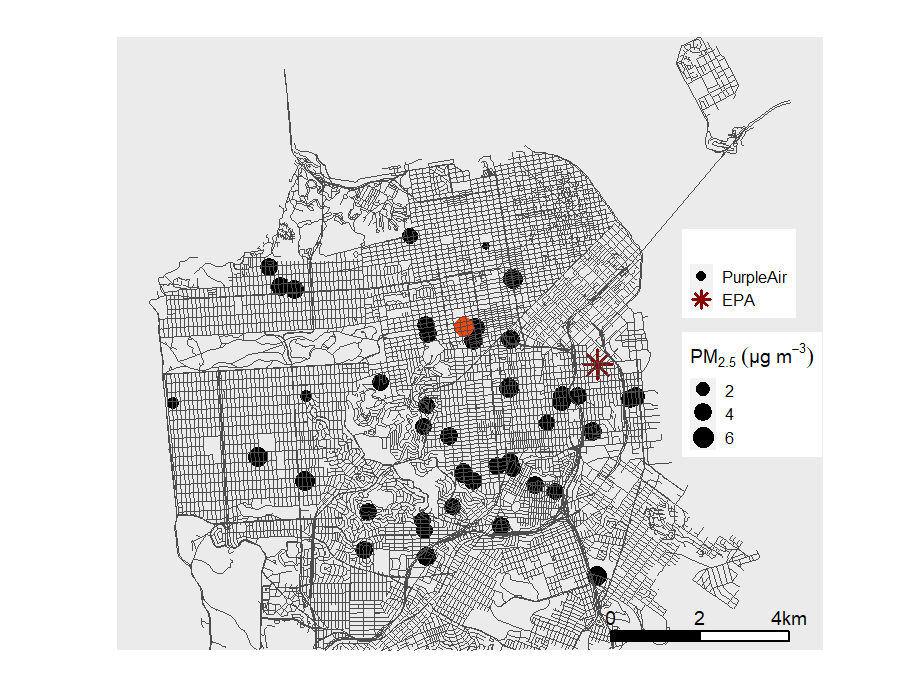}
\caption{Bubble plot of the median PM$_{2.5}$ concentration across the PurpleAir sensor locations (dots) and EPA site (cross) from January 1$^{\text{st}}$, 2020 through May 24$^{\text{th}}$, 2020. The orange location is the site depicted in Figure \ref{fig:AirPollAdjustPI}.}
\label{fig:AirPollLocs}
\end{figure}

We consider data from 44 different sensors located in San Francisco, CA, collected at hourly resolution but post-processed at 4-hour resolution forming a continuous record from 2020/01/01 to 2020/05/24, for a total of 870 time points. More than 100 additional PurpleAir stations were not considered because they had only a fraction of the data available from the study period. San Francisco was chosen due to its high coverage compared to other major US cities, as well as the availability of previous studies for validation of the sensors, as detailed below. The period of study was chosen to avoid the wildfire season in California \citep{mil12} with May 24, 2020 being three days before the first major seasonal wildfire \citep{cal21}, so that the local urban emissions are the main source of air pollution. Figure \ref{fig:AirPollLocs} shows the locations of the PurpleAir sites as well as the EPA's air quality monitoring station, along with the median measurement of the PM$_{2.5}$ concentration during the period of study. Previous work has validated the measurements from the PurpleAir sensors \citep{kelly20, ardon20, she21}. \cite{ardon20} showed that the sensors were generally well-correlated with the EPA observations, with the $R^2$ greater than 0.65 for 6 co-located sites in the United States, including San Francisco. In this work, the PurpleAir measurements were compared to the EPA data with a simple linear regression and achieved an $R^2$ of 0.52, thus in line with the previous study. A time series plot from the EPA site and the nearest PurpleAir site is shown in Figure S1A and a scatterplot with a fitted regression line can be seen in Figure S1B. In order to provide population exposure maps, population data from 194 census tracts in San Francisco was gathered from the 2014-2018 5-year American Community Survey by the US Census Bureau. 

\section{Methodology}\label{sec:method}

We introduce a deep ESN (DESN) in Section \ref{sec:DESN}, and the inference in Section \ref{sec:Infer} and we propose an adaptive method to quantify and adjust the forecast's uncertainty as well as its multivariate (possibly spatial) dependence in Section \ref{sec:Uncertainty}. 

\subsection{Deep Echo-State Networks}\label{sec:DESN}

A DESN is formulated as follows \citep{mcd18}:
\begin{subequations}\label{eqn:DESN}
\begin{flalign}
\text{output:} & \qquad \mb{Y}_{t} = \mb{B}_{D}\mb{h}_{t, D} + \sum_{d=1}^{D-1} \mb{B}_{d}k\left(\Tilde{\mb{h}}_{t,d}\right) + \bs{\epsilon}_{t} \label{eqn:DESN1}, \\
\text{hidden state $d$:} & \qquad \mb{h}_{t,d} = (1-\alpha)\mb{h}_{t-1,d} + \alpha\bs{\omega}_{t,d} \label{eqn:DESN2},\\
& \qquad \bs{\omega}_{t,d} = f_{h}\left(\frac{\nu_{d}}{|\lambda_{W_{d}}|} \mb{W}_{d} \mb{h}_{t-1,d} + \mb{W}_{d}^{\text{in}}\Tilde{\mb{h}}_{t,d-1}\right), \text{ for } d > 1 \label{eqn:DESN3},\\
\text{reduction $d-1$:} & \qquad \Tilde{\mb{h}}_{t,d-1} \equiv Q\left(\mb{h}_{t,d-1}\right), \text{ for } d > 1 \label{eqn:DESN4},\\
\text{input:} & \qquad \bs{\omega}_{t,1} = f_{h}\left(\frac{\nu_{1}}{|\lambda_{W_{1}}|} \mb{W}_{1} \mb{h}_{t-1, 1} + \mb{W}_{1}^{\text{in}}\mb{x}_{t}\right) \label{eqn:DESN5},\\
\text{matrix distribution:} & \qquad W_{d_{i,j}} = \gamma^{W_{d}}_{i,j} g(\eta_{W_{d}}) + (1-\gamma^{W_{d}}_{i,j})\delta_{0} \label{eqn:DESN6},\\
& \qquad W^{\text{in}}_{d_{i,j}} = \gamma^{W_{d}^{\text{in}}}_{i,j} g(\eta_{W_{d}^{\text{in}}}) + (1-\gamma^{W_{d}^{\text{in}}}_{i,j})\delta_{0} \label{eqn:DESN7},\\
& \qquad \gamma^{W_{d}}_{i,j} \text{\;$\thicksim$\;} Bern(\pi_{W_{d}}), \quad \gamma^{W_{d}^{\text{in}}}_{i,j} \text{\;$\thicksim$\;} Bern(\pi_{W_{d}^{\text{in}}}). \nonumber
\end{flalign}
\end{subequations}

The output $\mb{Y}_{t}$ represents the $n_{\ell}$-dimensional vector of air pollution (measured in PM$_{2.5}$ concentration) across all locations in San Francisco for time $t = 1, \ldots, T$ and is expressed as a linear function $\mb{B}_{D}\mb{h}_{t, D} + \sum_{d=1}^{D-1} \mb{B}_{d}k_{h}\left(\Tilde{\mb{h}}_{t,d}\right) + \bs{\epsilon}_{t}$ in \eqref{eqn:DESN1}, where $D$ represents the final hidden layer and the total number of layers in the DESN, $\mb{h}_{t,D}$ is an $n_{h,D}$-dimensional state vector, $\mb{\Tilde{h}}_{t,d}$ is an $n_{\Tilde{h},d}$-dimensional state vector, $\bs{\epsilon}_{t}$ is an error term, and the $\mb{B}$ are unknown coefficient matrices which are estimated via ridge regression with penalty $\lambda_{r}$. Here, $k(\cdot)$ in equation \eqref{eqn:DESN1} represents an activation function to ensure that $\Tilde{\mb{h}}_{t,d}$ is on a similar scale to $\mb{h}_{t,D}$. The state vector is a convex combination of a past state $\mb{h}_{t-1,d}$ and a memory term $\bs{\omega}_{t,d}$, controlled by a `leaking rate' parameter $\alpha$ as specified in \eqref{eqn:DESN2}. According to \eqref{eqn:DESN3}, the memory term $\bs{\omega}_{t,d}$ is the result of a nonlinear function $f_{h}$ with some pre-specified forms such as the hyperbolic tangent or rectified linear \citep{goo16}, which combines the past state $\mb{h}_{t-1,d}$ and some layer specific input data specified as either the $n_{x}$-dimensional vector of input $\mb{x}_{t}$, which in our application comprises of the past air pollution concentrations, i.e.,  $\mb{x}_{t}=\left(\mb{Y}_{t-\tau},\ldots,\mb{Y}_{t-m\tau}\right)^\top$, or the dimension reduced hidden state from the prior layer $\Tilde{\mb{h}}_{t,d-1}$. This ensures that at the present time, there is always a term accounting for the past air pollution up to a past time $t-m\tau$. The ability of the model to retain long-range information is controlled by parameters $\alpha$ and $m$, since the input comprises of values at time points $t-\tau$ through $t-m\tau$ where $\tau$ is commonly denoted as the forecast lead time \citep{mcd17}. Additionally, the function $Q(\cdot)$ in equation \eqref{eqn:DESN4} represents a dimension reduction function of the matrix $\mb{h}_{1:T,d}$ such that the dimension of the reduced matrix $\Tilde{\mb{h}}_{1:T,d}$ is less than if not equal to that of $\mb{h}_{1:T,d}$ (i.e., $n_{h,d} \ge n_{\Tilde{h}, d}$). For this work, we use an empirical orthogonal function (EOF) approach to act as $Q(\cdot)$ in the dimension reduction stage. Both matrices $\mb{W}_{d}$ and $\mb{W}_{d}^{\text{in}}$ in \eqref{eqn:DESN3} are randomly generated assuming high sparsity. Indeed, the entries of both matrices are a linear combination of $\delta_{0}$, a Dirac function at zero, and a symmetric distribution $g(\cdot)$ centered about zero with hyper-parameters $\eta_{W_{d}}$ and $\eta_{W_{d}^{\text{in}}}$, see \eqref{eqn:DESN4} and \eqref{eqn:DESN5}. The distribution of each entry is therefore a `spike-and-slab', a widespread choice for prior in Bayesian variable selection \citep{ishwa05, malsi11}. The probability of each entry being zero is controlled by independent Bernoulli random variables with probabilities $\pi_{W_{d}}$ and $\pi_{W_{d}^{\text{in}}}$ for $\mb{W}_{d}$ and $\mb{W}_{d}^{\text{in}}$, respectively The error $\bs{\epsilon}_{t}$ is independent identically distributed mean zero multivariate normal with some covariance structure. In this work, instead of specifying it explicitly, we account for the dependence after the forecasts have been produced, as we will show in Section \ref{sec:Uncertainty}.

In order to be well defined, the proposed ESN must also have the \textit{echo-state property}, i.e., with a sufficiently long time sequence the model must asymptotically lose its dependence on the initial conditions \citep{luk12,jae07}. In order for this property to hold, the spectral radius or largest eigenvalue of $\mb{W_{d}}$, denoted by $\lambda_{W_{d}}$, must be less than one. This is ensured by the use of the scaling parameter $\nu_{d}$. The set of hyper-parameters for this model is then $\bs{\theta}=(n_{h,d}, n_{\Tilde{h}, d}, m, \nu_{d},\lambda_{r}, \eta_{W_{d}}, \eta_{W_{d}^{\text{in}}}, \pi_{W_{d}}, \pi_{W_{d}^{\text{in}}}, \alpha, D)$, where hyper-parameters denoted with $d$ are to be layer specific choices.

\subsection{Inference and Computational Issues}\label{sec:Infer}

In order to perform inference on the hyper-parameter vector $\bs{\theta}$, a validation approach is used, where a portion of the training data is held out (validation set) and forecasts are validated over it. The distribution $g(\cdot)$ used to populate the matrices $\mb{W}$ and $\mb{W}^{\text{in}}$ is typically generated from either a symmetric uniform or a normal distribution centered around zero \citep{luk12}. In this work we use standard normal distributions to populate these matrices, thus we remove hyper-parameters $\eta_{W_{d}}$ and $\eta_{W_{d}^{\text{in}}}$. \cite{mcd18} found that fixing the  hyper-parameters $\pi_{W_{d}}$ and $\pi_{W_{d}^{\text{in}}}$ across each layer was a reasonable assumption, thus we fix these values to be 0.1. Additionally, the prior work showed that fixing $n_{h,d}$ and $n_{\Tilde{h},d}$ in all layers except the last layer $D$ ultimately improved forecasting capabilities, thus for simplicity we shall denote $n_{h,d}$ and $n_{\Tilde{h},d}$ as $n_{h}$ and $n_{\Tilde{h}}$, respectively. Hence, the hyper-parameter space is reduced to $\bs{\theta}^{*}=(n_{h}, n_{h,D}, n_{\Tilde{h}}, m, \nu_{d}, \lambda_{r}, \alpha, D)$. For any given $\bs{\theta}^*$ the matrix $\mathbf{B}$ in \eqref{eqn:DESN1} can be computed. This enables to produce predictions in the validation set, and hence to compute the mean squared error (MSE). Inference on $\bs{\theta}^*$ is then performed by finding the minimizer of this value.

A (conditional) regularized regression such as ridge regression with penalty $\lambda_{r}$ mitigates the large variance of the estimators. Equation \eqref{eqn:DESN1} can be denoted in matrix form as $\mb{Y} = \mb{HB} + \mb{E}$, where
\begin{gather}
 \mb{Y} = \begin{bmatrix} \mb{Y}^{\top}_{1} \\ \vdots \\ \mb{Y}^{\top}_{T} \end{bmatrix}, \quad
 \mb{H} = \begin{bmatrix} \mb{h}^{\top}_{1} \\ \vdots \\  \mb{h}^{\top}_{T} \end{bmatrix},  \quad
 \textbf{E} = \begin{bmatrix} \mbox{\boldmath{$\epsilon$}}^{\top}_{1} \\ \vdots \\ \mbox{\boldmath{$\epsilon$}}^{\top}_{T}. \end{bmatrix},
 \nonumber
\end{gather}

\noindent and $\mb{h}^{\top}_{t} = \left(\mb{h}_{t,D}, k(\mb{\Tilde{h}}_{t,1}), \dots,  k(\mb{\Tilde{h}}_{t,D-1})\right)^{\top}$, for $t=1,\dots,T$. A ridge regression is applied to estimate $\mb{B}$ such that $\hat{\mb{B}} = \argmin_{\mb{B}}(\|\mb{Y} - \mb{HB}\|^{2}_{F} + \lambda_{r}\|\mb{B}\|^{2}_{F})$, where $\|\cdot\|_{F}$ is the Frobenius norm and $\lambda_{r}$ is the penalty parameter. This provides the ridge estimator $\hat{\mb{B}}$ in closed form as $\hat{\mb{B}} = (\textbf{H}^{\top}\mb{H} + \lambda_{r}\mb{I})^{-1}\mb{H}^{\top}\mb{Y}$.

Additionally, while the sparsity of the weight matrices allows for fast and memory efficient sparse linear algebra operations, the scalability of the inference and hence the generation of the ensemble for many locations in $\mb{Y}_{t}$ represents a challenge, especially in cases where the time scale is fine (e.g., $<$1 day) and the forecasts need to be provided at most at a fraction of the sampling rate. Generating ensemble forecasts independently in parallel greatly improves the computational efficiency, as each simulation requires linear algebra operations with high-dimensional output vectors, representing spatial locations in a typical spatio-temporal application. In the supplementary material we assess the computational speedup as a function of the number of processing units (cores) used to generate the ensembles of forecasts. 

\subsection{Calibrating the Forecast}\label{sec:Uncertainty}

One might attempt to quantify uncertainty at a future time point by feeding the previous ensemble estimates forward, thus creating an ensemble of ensembles.  However, for either long time series or high-dimensional data vectors, this approach would rapidly become computationally expensive and impractical. In this work, we propose a new method to quantify the forecast uncertainty which is feasible both in terms of time and memory. The method relies on the use of a fraction of the training set to 1) compare predictions and true values 2) calibrate the marginal uncertainty of the residuals and 3) estimate their correlation. Throughout this section we assume that data are observed for $t=1,\ldots, T+n_{w}\tau$ but the model is only trained until $T$. We generate an ensemble of forecasts for time points $T+1$ through $T+n_{w}\tau$ across multiple `windows', for some long-lead forecast of $\tau$ steps ahead, which are then used for calibration. More specifically, ensembles of forecasts would be generated for the time points $T+1, \dots, T+n_{w} \tau$ in intervals of size $\tau$ (representing the forecast horizon), where $n_{w}$ represents the number of `windows'. In order to obtain a robust quantification of the uncertainty it is recommended that the maximum amount of windows be chosen such that the forecasting ability on earlier windows is not sacrificed due to lack of available training data. We denote the forecast for time point $T+1$ given $\{\mb{Y}_{T}, \mb{Y}_{T-1}, \dots\}$ as $\hat{\mb{Y}}^{T}_{T+1} = \EX(\mb{Y}_{T+1}|\mb{Y}_{T}, \mb{Y}_{T-1}, \dots)$, where the expectation is intended with respect to model \eqref{eqn:DESN}. The forecasts are then grouped according to each vector element $\ell=1,\dots,n_{\ell}$ as follows:
\begin{gather}\label{eqn:forcgroups}
 \bs{f}^{(\ell)}_{1} = \begin{bmatrix} \hat{Y}_{T+1}^{T}(\ell) \\ \hat{Y}_{T+\tau+1}^{T+\tau}(\ell) \\ \vdots \\ \hat{Y}^{T+(n_{w}-1) \tau}_{T+(n_{w}-1) \tau+1}(\ell)  \end{bmatrix}, \hspace{0.20em} \bs{f}^{(\ell)}_{2} = \begin{bmatrix} \hat{Y}_{T+2}^{T}(\ell) \\ \hat{Y}^{T+\tau}_{T+\tau+2}(\ell) \\ \vdots \\ \hat{Y}^{T+(n_{w}-1) \tau}_{T+(n_{w}-1) \tau+2}(\ell)  \end{bmatrix}, \dots, \hspace{0.25em} \bs{f}^{(\ell)}_{\tau} = \begin{bmatrix} \hat{Y}^{T}_{T+\tau}(\ell) \\ \hat{Y}^{T+\tau}_{T+2 \tau}(\ell) \\ \vdots \\ \hat{Y}^{T+(n_{w}-1) \tau}_{T+n_{w}  \tau}(\ell)  \end{bmatrix}.
\end{gather}
Since these groupings are created for each vector element independently, the length of each vector $\left\{\bs{f}^{(\ell)}_{j}; j = 1, \dots, \tau\right\}$ is $n_{w}$. As an example, if $\tau=20$ and $n_{w} = 10$ then the one- and two-step ahead forecast for vector element $\ell$ would be grouped as $\bs{f}^{(\ell)}_{1} = \left(\hat{Y}^{T}_{T+1}(\ell), \hat{Y}^{T+20}_{T+21}(\ell), \dots, \hat{Y}^{T+180}_{T+181}(\ell)\right)$ and $\bs{f}^{(\ell)}_{2} = \left(\hat{Y}^{T}_{T+2}(\ell), \hat{Y}^{T+20}_{T+22}(\ell), \dots, \hat{Y}^{T+180}_{T+182}(\ell)\right)$. The residuals associated with each of these groupings are calculated by subtracting the forecast from each associated true value, e.g., for the $j$-step ahead forecasts and vector element $\ell$ we have:
\begin{flalign}\label{eqn:residcalc}
& \mb{R}^{(\ell)}_{j} = \mb{Y}^{(\ell)}_{j\text{-step}}-\mb{f}^{(\ell)}_{j}, \quad j=1,\dots,\tau,
\end{flalign}
where $\mb{Y}^{(\ell)}_{j\text{-step}} = \{Y_{T+(w-1) \tau+j}(\ell); w=1,\dots, n_{w}\}$ and $\mb{R}^{(\ell)}_{j} = \{R^{(\ell)}_{j}(w); w=1,\dots, n_{w}\}$ . The standard deviation associated with the residuals of each collection of future time points is then calculated as:
\begin{flalign}\label{eqn:SDCalc}
& \hat{{\sigma}}^{(\ell)}_{j} = \sqrt{\frac{\sum^{n_{w}}_{w=1}\left({R}^{(\ell)}_{j}(w)-\overline{{R}^{(\ell)}_{j}}\right)^2}{n_{w}-1}}, \quad j=1,\dots,\tau,
\end{flalign}
where $\overline{{R}^{(\ell)}_{j}}=\frac{\sum^{n_{w}}_{w=1}{R}^{(\ell)}_{j}(w)}{n_{w}}$. A reasonable assumption is that uncertainty about the forecasts for each vector element expands in time, i.e., ${\hat{\sigma}}^{(\ell)}_{j} \leq {\hat{\sigma}}^{(\ell)}_{j^{'}}, j^{'} \ge j$ for every $\ell$. The vector $\left({\hat{\sigma}}^{(\ell)}_{1},\dots,{\hat{\sigma}}^{(\ell)}_{\tau}\right)$ is then smoothed with a monotonic increasing spline with resulting fitted values indicated as $\left(\tilde{{\sigma}}^{(\ell)}_{1},\dots,\tilde{{\sigma}}^{(\ell)}_{\tau}\right)$. In this work, we use a monotonic cubic spline interpolation \citep{Wolberg99, deBoor78} with the additional constraint $\tilde{\sigma}^{\prime(\ell)}_{j} \geq 0$, where $\tilde{\sigma}^{\prime(\ell)}_{j}$ represents the slope of the spline at future point $j$. This method would also allow for calibration of the forecasts \citep{gneit07}, as the ensemble is adjusted to quantify the correct uncertainty, as will be shown by the probability integral transform (PIT) of the standardized residuals:
\begin{flalign}\label{eqn:StandResids}
\widetilde{\mb{R}}^{(\ell)}_{j}=\{\mb{R}^{(\ell)}_{j}/\tilde{{\sigma}}^{(\ell)}_{j}; j=1,\dots,\tau\}. 
\end{flalign}
Since the residuals are assumed to follow a multivariate Gaussian distribution by the model in equation \eqref{eqn:DESN} and the forecasts are calibrated, the PIT will more closely resemble a standard uniform distribution after the proposed adjustment and under the assumption that training and validation data have the same behavior, the proposed approach leads to consistent predictions \citep{hai18}. Additionally, the estimated marginal standard deviations from equation \eqref{eqn:SDCalc} can then be used to generate the confidence intervals for each vector element, as will be shown in Sections \ref{sec:sim} and \ref{sec:appl}. The full approach is detailed in Algorithm \ref{alg:UncAdj}.
\begin{algorithm}[!tb]
\caption{Marginal forecast calibration.}
\label{alg:UncAdj}
\SetAlgoLined
\KwData{$\textbf{Y}_{t}: t = 1, \dots, T+n_{w}\tau$}
\KwResult{Vector of standard deviations $\tilde{\bs{\sigma}}$}
\textbf{Initialize:} Generate forecasts $\hat{\mb{Y}}^{(\ell)}_{j\text{-step}} = \left\{\hat{Y}^{T+(w-1)\tau}_{T+(w-1) \tau+j}(\ell); \hspace{0.25em} w=1,\dots, n_{w}\right\}$ for $j=1, \dots, \tau$ and $\ell = 1, \dots, n_{\ell}$\\
 \For{$\ell = 1, \dots, n_{\ell}$}{
    \For{$j=1, \dots, \tau$}{
         Compute the $j$-step ahead forecast $\left\{\hat{Y}^{T+(w-1)\tau}_{T+(w-1)\tau+j}(\ell); \hspace{0.25em} w=1,\dots, n_{w}\right\}$ and group them into $\mb{f}^{(\ell)}_{j}$, as detailed in equation \eqref{eqn:forcgroups}\;
         Calculate the residuals for the $j$-step ahead forecast $\mb{R}^{(\ell)}_{j}$ from equation \eqref{eqn:residcalc}\; 
        Compute the $j$-step ahead residual standard deviation $\hat{{\sigma}}^{(\ell)}_{j}$ from equation \eqref{eqn:SDCalc}\;
    }
    Fit a monotonic increasing spline to  \{$\hat{{\sigma}}^{(\ell)}_{j};\quad j=1,\dots,\tau$\} and denote the vector of fitted values $\tilde{\bs{\sigma}}$.
 }
\end{algorithm}

Finally, the vector of standardized residuals $\widetilde{\mb{R}}_{j,w}=\left(\widetilde{R}^{(1)}_{j}(w),\ldots, \widetilde{R}^{(n_\ell)}_{j}(w)\right)$ is assumed to have a multivariate $n_{\ell}$-dimensional normal distribution across vector elements represented as:
\begin{flalign}\label{eqn:multinorm}
& \widetilde{\mb{R}}_{j,w} \stackrel{i.i.d.}{\sim} N_{n_{\ell}}(\bs{0}, \textbf{C}),
\end{flalign}
i.e., the $\widetilde{\mb{R}}_{j,w}$ are independent and identically distributed across $j=1,\ldots, \tau$ and $w=1,\ldots, n_w$ with correlation matrix $\textbf{C}$. In this work, we use a penalized non-parametric method to generate sparse estimates of \textbf{C} in the simulation study in Section \ref{sec:sim}, and a parametric model for the spatial application in Section \ref{sec:appl}.

\section{Simulation Study}\label{sec:sim}

 In this work, we aim to understand the predictive capabilities of the DESN on time series with non-linear dynamics against more traditional time series models. We do this in a controlled setting from a model with a well known quasi-periodic behavior. In Section \ref{sec:Lor96Data} we simulate data from a Lorenz-96 model \citep{lor96}, a model first proposed to study forecasting ability on spatial chaotic systems \citep{Kar10}. We compare the ESN with other standard time series methods in terms of their predictive ability in Section \ref{sec:Lor96Forcs}. Finally, we calibrate the marginal forecasts, estimate the multivariate dependence and compare the results with both an uncalibrated and independent approach in Section \ref{sec:Lor96Uncertain}.

\subsection{Simulated Data}\label{sec:Lor96Data}
 We simulate data from the \cite{lor96} differential equations for $n_{\ell}=40$ variables, i.e.,
\begin{eqnarray}\label{eqn:lor96funcs}
\frac{\mathrm{d}Y_{t}(\ell)}{\mathrm{d}t} = \left\{Y_{t}(\ell+1) - Y_{t}(\ell-2)\right\}Y_{t}(\ell-1) - Y_{t}(\ell) + F_{t}, \quad \ell = 1, \dots, 40.
\end{eqnarray}
The forcing is constant in time and set to $F_{t} = 4.5$, which implies a median absolute correlation of 0.42 amongst the $\ell$ vector elements. In the supplementary material, we show an analysis with different values of $F_t$, implying a varying degree of dependence among vector elements. Additionally, we simulated ten realizations of this data from ten different random initial conditions to be used for this analysis. A realization of 200 time points from a selected vector element for one out of the ten realizations is shown in Figure S5. The ESN was initially trained using the first $T=980$ time points and was tested using the subsequent $\tau=20$ time points.  The grids used for hyper-parameter selection are as follows: $n_{h, d}=$\{$50, 150, 500, 750$\}, $n_{h,D}=$\{$40, 45, 50$\}, $n_{\Tilde{h}}=$\{$10, 15, 20, 25$\}, $m = $\{$1, \dots, 4$\}, $\nu_{d} = $\{$0.2 \times s_{\nu_{d}} : s_{\nu_{d}} = 1, \dots, 5$\}, $\lambda_{r}$ = \{$0.001,0.01,0.1$\}, $\alpha = $\{$0.01 \times s_{\alpha} : s_{\alpha} = 1, \dots, 100$\}, and we chose $D = \{2,3,4\}$ layers. After validation, we found the following hyper-parameters for the DESN to be optimal: $D=3$, $n_{h,d} = 500$, $n_{h,D} = 40$, $n_{\Tilde{h}} = 10$, $\lambda_{r} = 0.1$, $\nu_{d} = \{0.4, 1.0, 1.0\}$, $m = 1$, $\alpha = 0.49$, $\pi_{W} = \pi_{W^{\text{in}}} = 0.1$ and $g(\cdot)$ set to a standard normal distribution. Instead of focusing on short-range forecasts as in \cite{mcd17} where $\alpha = 1$, however, this work focuses on the ability of the ESN model to produce accurate long-range forecasts which is highly dependent on the memory hyper-parameter $\alpha$ which is conveyed through the choice of a smaller $\alpha=0.49$ value.

\subsection{Point Forecasts}\label{sec:Lor96Forcs}
\quad In order to compare the ESN with traditional time series modelling approaches, we also fit an auto-regressive fractionally integrated moving average model (ARFIMA; \cite{granger80, hosking81}). ARFIMA$(p,d,q)$ can be written in operator form as
\begin{eqnarray}\label{eqn:ARIMA}
\left(1 - \sum_{i=1}^{p} \phi_{i}B^{i}\right)(1-B)^{d}X_{t} = \left(1 + \sum_{i=1}^{q} \theta_{i}B^{i}\right)\epsilon_{t},
\end{eqnarray}
where $B$  is the backshift operator: $BX_{t}=X_{t-1}$ and $\epsilon_{t}\stackrel{i.i.d.}{\sim}N(0,\sigma^2)$, i.e., independent and identically distributed zero mean Gaussian random variables. ARFIMA models are a generalized version of the popular auto-regressive integrated moving average (ARIMA; \cite{brockwell16}) model with non-integer difference parameter $d$. Additionally, we consider a multivariate version of this model called vector autoregression (VAR, \cite{hyn21}), defined as:
\begin{gather}
 \begin{bmatrix} {Y}_{t}(1) \\ \vdots \\ {Y}_{t}(n_{\ell}) \end{bmatrix} = \mb{A}_{0} + \mb{A}_{1}\begin{bmatrix} {Y}_{t-1}(1) \\ \vdots \\ {Y}_{t-1}(n_{\ell}) \end{bmatrix} + \dots + \mb{A}_{p}\begin{bmatrix} {Y}_{t-p}(1) \\ \vdots \\ {Y}_{t-p}(n_{\ell}) \end{bmatrix} + \begin{bmatrix} \mbox{$\epsilon$}_{t}(1) \\ \vdots \\ \mbox{$\epsilon$}_{t}(n_{\ell}). \end{bmatrix},
 \nonumber
\end{gather}

\noindent where $p$ is the order of the autoregression term, $\mb{A}_{0}$ is a vector of unknown parameters, $\mb{A}_{1}, \dots, \mb{A}_{p}$ are unknown matrices, $\epsilon_{t}(1), \ldots, \epsilon_{t}(n_{\ell})$ are error terms. Unlike ARIMA, these VAR models can generate forecasts for all vector elements simultaneously.

We also fit a state-space model \citep{durbin12}. The state-space model assumes the `state variables' at time $t$ are derived from the state at time $t-1$ and that the output variables are dependent upon the value of the state variables at time $t$. In this work, we treat the ESN as a special case of the state-space model by assuming that $\alpha=1$ in \eqref{eqn:DESN2} and $f_h(\cdot)$ is the identity function in \eqref{eqn:DESN3}. The ESN in equation \eqref{eqn:DESN} would then take the form:
\begin{subequations}\label{eqn:StateSpace}
\begin{flalign}
& \bs{Y}_{t} = \mb{B}\bs{h}_{t} + \bs{\epsilon}_{t},\label{eqn:modKalman1}\\
& \bs{h}_{t} = \frac{\nu}{|\lambda_{W}|} \bs{W} \bs{h}_{t-1} + \bs{W}^{\text{in}}\bs{x}_{t}.\label{eqn:modKalman2}
\end{flalign}
\end{subequations}

We also consider a hidden Markov model (HMM, \cite{rab89}) as a comparison. HMMs are similar to state-space models where the system being observed is modelled via unobserved hidden states, but have the additional assumption that the system is a Markov process, i.e., the data is dependent on past observations up until some points in the past, and then is conditionally independent. Here we consider a HMM as a specific case of the state-space model where $\bs{x}_{t} = \bs{Y}_{t-1}$ and the scaled weight matrix $\frac{\nu}{|\lambda_{W}|}\bs{W}$ is the transition matrix. These models have been shown previously to accommodate a large variety of temporal behaviors in applications ranging from speech recognition \citep{rab89} to electrical engineering \citep{sat93}.

Finally, we also consider two reference machine learning methods for multivariate time series for comparison: a long short-term memory (LSTM, \cite{hoch97}) model and a gated recurrent unit (GRU, \cite{cho14}) model. These models have similar structure to that of the ESN but have a considerably large parametric space as they assume fully connected, dense, weight matrices and are optimized through gradient-based methods (backpropagation). To ensure a fair comparison, the number of hidden nodes in the LSTM and GRU ($n_{h}=500$) and the number of past time points used as input data ($m=1$) is identical. A further discussion on these models is deferred to the supplementary.

\begin{figure}[!tb]
\centering
\includegraphics[width = 10cm]{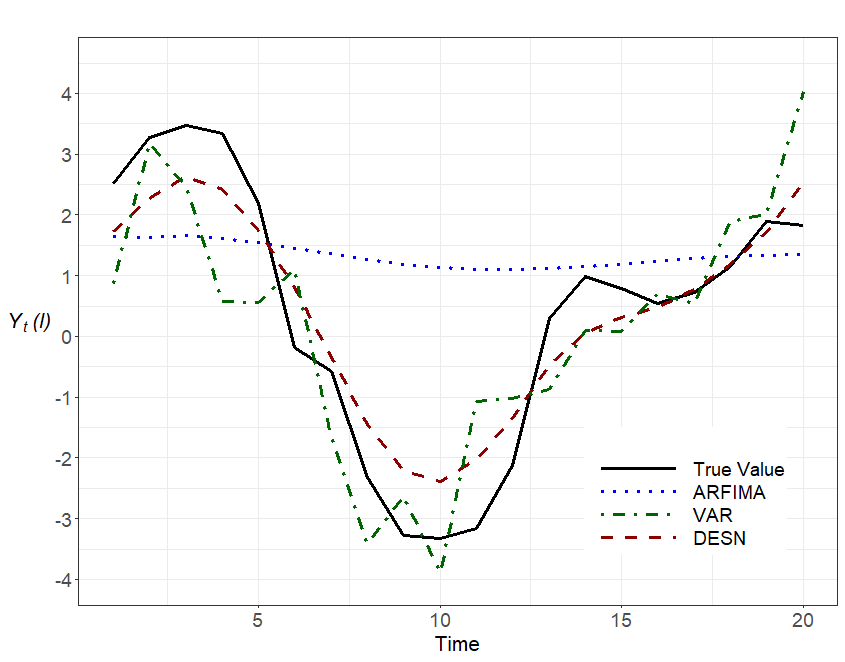}
\caption{Comparison of different methods for the same vector element from one ensemble member of the \cite{lor96} data with long-range forecasts for $\tau=20$ time points}
\label{fig:L96Comparison}
\end{figure}

The forecasts from the DESN, ARFIMA, and VAR models are shown in Figure \ref{fig:L96Comparison}, and it can seen that the ARFIMA and VAR models produce inadequate forecasts compared to the DESN. The sub-optimal performance can be attributed to the underlying nonlinear dynamics of the \cite{lor96} model. For example, ARIMA models are designed to only capture linear dependence, and are not sufficiently flexible to capture a highly nonlinear behavior such as that of \cite{lor96}. We quantify the forecasting performance for all of the different methods using the MSE and the continuous ranked probability score (CPRS; \cite{gneit07}), a proper scoring rule for probabilistic forecasts which would be equal to zero with a perfect forecast. The results, shown in Table \ref{tbl:MethodComp}, highlight a median MSE across vector elements and realizations of 2.92 with interquartile range (IQR) of 2.16 for the ARFIMA model, whereas the state-space model returns an MSE of 4.58 (2.75). The HMM generates an MSE of 9.92 (11.37), the VAR model returns an MSE of 1.62 (1.09) and the LSTM and GRU models generate MSE values of 6.00 (6.39) and 6.67 (7.11), respectively. The suboptimal performance of the state-space and HMM models is attributable to the lack of long-range dependence being captured by these models, while LSTM and GRU's performance can be largely attributed to the lack of a sufficiently long record of training data available to train these models, whose large parametric space require a substantial amount of information for proper training. The DESN clearly outperforms all of these models with MSE of 0.66 (0.63), while a shallow version ($D=1$) of the ESN model would lead to an MSE of 1.26 (0.90). The same relative ordering of forecasting performance was found for the CRPS, with the DESN model showing the best performance. 

\begin{table}[!tb]
\begin{centering}
\begin{tabular}{||c|c|c||} 
\hline
Forecasting Method & MSE & CRPS \\ [1ex] 
\hline\hline
Deep ESN & 0.66 (0.63) & 0.61 (0.15)  \\
\hline
Shallow ESN & 1.26 (0.90) & 0.72 (0.20)  \\
\hline
ARFIMA & 2.90 (2.19) & 0.98 (0.36)  \\
\hline
State-Space & 4.58 (2.75) & 1.23 (0.37)  \\ 
\hline
Hidden Markov Model & 9.92 (11.37) & 1.99 (1.22) \\
\hline
Vector Autoregression & 1.74 (1.11) & 0.80 (0.21) \\ 
\hline
Long Short-Term Memory & 6.00 (6.39) & 1.46 (0.83) \\
\hline
Gated Recurrent Unit & 6.67 (7.11) & 1.54 (0.93) \\[1ex]
\hline
\end{tabular}
\caption{Comparison of the forecasting methods in terms of MSE and CRPS across all vector elements from each of the 10 realizations of simulated \cite{lor96} data in equation \eqref{eqn:lor96funcs}. A median across all vector elements and realizations is shown, and in parenthesis the IQR is reported.}
\label{tbl:MethodComp}
\end{centering}
\end{table}

\subsection{Prediction Uncertainty}\label{sec:Lor96Uncertain}

\begin{figure}[!tb]
\centering
\includegraphics[width = 13cm]{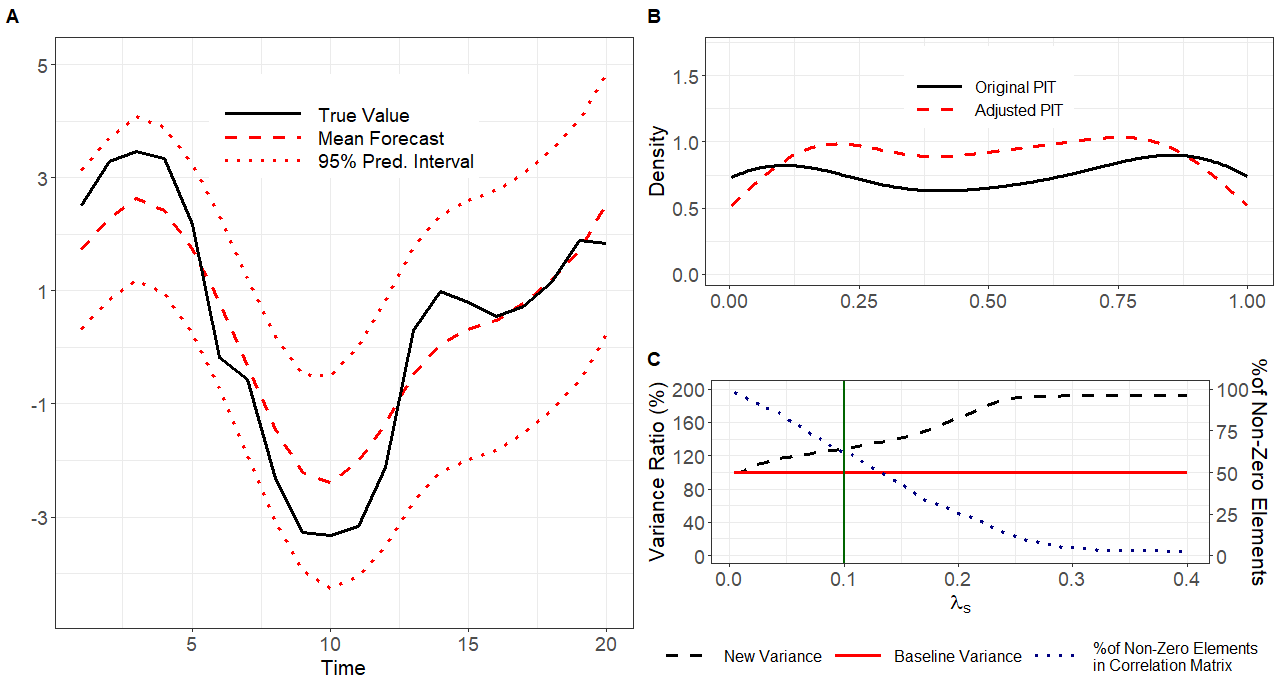}
\caption{Uncertainty quantification for the same vector element from Figure \ref{fig:L96Comparison}. (A) Forecasts with adjusted PIs generated using standard deviations from equation \eqref{eqn:SDCalc}, (B) original and adjusted PIT for the standardized residuals from the forecasts and (C) percentage ratio between the new and baseline variance of the mean standardized residuals and proportion of non-zero elements in correlation matrix as the penalty parameter $\lambda_{s}$ increases in equation \eqref{eqn:SparseCorrel}. The green line represents the $\lambda_{s}=0.1$ value chosen to generate the sparse correlation matrix heatmap in Figure S6B.}
\label{fig:Lor96AdjustPI}
\end{figure}

The calibration method discussed in Section \ref{sec:Uncertainty} was implemented with the ESN for $n_{w}=20, \tau=20$ for all 10 realizations from the Lorenz 96 equation \eqref{eqn:lor96funcs}. The first row of Table \ref{tbl:PICoverage} shows that, when using the uncalibrated ensemble forecasts, the marginal uncertainty is incorrectly quantified and leads to a considerably smaller coverage for the nominal 95\%, 80\%, and 60\% prediction intervals (PIs), with an average coverage discrepancy of 19.3\% across the three confidence levels. When using the adjusted standard deviation for the calibration approach, however, the marginal forecast uncertainty is properly quantified and does capture the appropriate amount of data within each of the specified PIs. Indeed, the calibrated 95\% prediction interval on row 2 captures a median of 94.2\% of the data across all vector elements with an IQR of 2.8\%, a figure comparable to the 1.2\% average coverage discrepancy across all three intervals. Figure \ref{fig:Lor96AdjustPI}A shows the calibrated 95\% PIs for a sample vector element where the increase in the uncertainty through time is readily apparent, due to the use of the monotonic spline on the standard deviations. Figure \ref{fig:Lor96AdjustPI}B shows instead how the calibrated PIT is closer to uniform than the uncalibrated PIT thereby underscoring an improved quantification of the uncertainty, as expected from the discussion in Section \ref{sec:Uncertainty}.
\begin{table}[tb!]
\begin{centering}
\begin{tabular}{||c|c|c|c|c||} 
\hline
Coverage & Method & 95\% & 80\% & 60\% \\ [1ex] 
\hline\hline
\multirow{2}{*}{Marginal} & Uncalibrated  & 77.5 (4.7) & 58.3 (4.6) & 41.2 (3.9)\\
\cline{2-5}
& Calibrated & 94.2 (2.8) & 80.9 (4.7) & 62.0 (4.8)\\
\hline
\hline
\multirow{3}{*}{Difference} & Independent & 94.7 (2.3) & 82.4 (4.2) & 63.9 (3.9)\\
\cline{2-5}
& Dependent & 90.9 (4.8) & 75.5 (6.9) & 56.8 (8.0)\\
\cline{2-5}
& Dependent \& Sparse & 93.8 (1.7) & 80.5 (2.2) & 61.8 (3.0)\\
\hline
\hline
\multirow{3}{*}{Mean} & Independent & 99.9 (0.4) & 96.5 (0.9) & 82.5 (1.9)\\
\cline{2-5}
& Dependent & 91.6 (2.4) & 73.7 (3.1) & 54.3 (1.4)\\
\cline{2-5}
& Dependent \& Sparse & 99.0 (1.1) & 90.6 (2.4) & 72.7 (2.1)\\
\hline
\end{tabular}
\caption{Coverage from: marginal prediction intervals (rows 1-2), the difference between neighboring elements (rows 3-5), and the grand mean of forecasts (rows 6-8) across each of the 10 realizations of simulated \cite{lor96} data in equation \eqref{eqn:lor96funcs}. The median is shown and in parenthesis the IQR of the coverage is also reported.}
\label{tbl:PICoverage}
\end{centering}
\end{table}

In order to estimate the correlation matrix \textbf{C} of the standardized residuals in equation \eqref{eqn:multinorm}, we propose a penalized non-parametric sparse method. We use the majorize-minimize algorithm proposed by \cite{bien11} to estimate a sparse correlation matrix as determined by some pre-specified penalty parameter $\lambda_{s}$. The minimization problem among all symmetric non-negative matrices $\textbf{C}\ge0$ can be formulated as follows:
\begin{eqnarray}\label{eqn:SparseCorrel}
\hat{\textbf{C}}_{\text{sparse}}(\lambda_{s}) = \argmin_{\textbf{C}\ge0}\left[\log{\left\{\det{(\textbf{C})}\right\}} + \tr{(\textbf{C}^{-1}* \hat{\textbf{C}})} + \lambda_{s}||\textbf{P}*\textbf{C}||_{1}\right],\nonumber
\end{eqnarray}
where $\hat{\textbf{C}}$ is the sample correlation matrix of the standardized residuals of the adjusted forecasts in equation \eqref{eqn:StandResids}, \textbf{P} is a matrix including zeros on the diagonal and ones elsewhere, and $*$ indicates element-wise multiplication which ensures that only the off-diagonal elements are penalized. The sparsity can be controlled by evaluating the trade-off between the proportion of zeros in the matrix and loss of information in the form of changing variation of the standardized residuals. Figure \ref{fig:Lor96AdjustPI}C shows the change of the variance of the average standardized residuals, $\sigma^{2}_{\overline{\widetilde{R}}}(\lambda_{s}) = \frac{1}{{n_{\ell}}^2}\textbf{1}^{\top}\hat{\textbf{C}}_{\text{sparse}}(\lambda_{s})\textbf{1}$ (\textbf{1} is a $n_{\ell}$-dimensional column vector of ones), as well as the proportion of non-zero elements in $\hat{\textbf{C}}_{\text{sparse}}(\lambda_{s})$ with respect to the penalty parameter $\lambda_{s}$. Here, the red line represents the baseline variance assuming $\lambda_{s} = 0$, the dashed black line represents the percentage change of $\sigma^{2}_{\overline{\widetilde{R}}}(\lambda_{s})$ against $\sigma^{2}_{\overline{\widetilde{R}}}(0)$ as a function of $\lambda_{s}$, and the dotted blue line represents the proportion of non-zero elements. Heatmaps of $\hat{\textbf{C}}$ and $\hat{\textbf{C}}_{\text{sparse}}(\lambda_{s} = 0.1)$ (represented as the solid green line in Figure \ref{fig:Lor96AdjustPI}C) can be seen in panels A and B of Figure S6. 

$\hat{\textbf{C}}_{\text{sparse}}(\lambda_{s}=0.1)$ was then used to calculate the nominal coverage of PIs for $\mb{R}^{(\ell)}_{j} - \mb{R}^{(\ell+1)}_{j}, j=1,\dots,\tau$, the difference in residuals between neighboring vector elements. More specifically, their variance was calculated assuming both independence and dependence (from $\hat{\textbf{C}}$) among the elements. For this coverage calculation, only vector elements with mild absolute correlation between 0.4 and 0.6 were considered and this constraint was applied across all 10 realizations of the simulated data. It is shown in rows 3-5 of Table \ref{tbl:PICoverage} that the nominal coverage improves in the sparse-dependent case where the adjusted uncertainty is used to determine the dependence among the vector elements, for the average coverage discrepancy is 1.2\% versus 2.2\% shown in the independent case.

The median coverage of the grand mean, i.e., the average across all vector elements, is shown in rows 6-8 in Table \ref{tbl:PICoverage}. When the vector elements are treated independently for each realization, there is an average over-coverage of 14.6\% across the chosen intervals. The nominal coverage improves when $\hat{\textbf{C}}$ and $\hat{\textbf{C}}_{\text{sparse}}(\lambda_{s}=0.1)$ are used to inform dependence with an average under-coverage of 5.1\% and an average over-coverage of 9.1\% across the intervals, respectively. By sparsifying the empirical correlation matrix, the overall variance is reduced, therefore explaining the improved coverage versus the independent case. 

\section{Application}\label{sec:appl}

We now use the proposed approach to produce long-range forecasts with calibrated uncertainty quantification for the San Francisco air pollution data presented in Section \ref{sec:data}. The first $T=840$ points, consisting of measurements from 2020/01/01 to 2020/05/19, were used as a training set, while the remaining $\tau=30$ points, representing 2020/05/20 through 2020/05/24, were used as the testing set. The grids for all the hyper-parameters were the same as those used for the \cite{lor96} simulated data in Section \ref{sec:Lor96Forcs}. All of the hyper-parameter grids were optimized and evaluated using MSE across all time points for all locations using a forecast ensemble of 300 elements. The hyper-parameters which produced the minimum validation MSE were: $D=3$, $n_{h,d} = 500$, $n_{h,D} = 45$, $n_{\Tilde{h}} = 20$, $\lambda_{r} = 0.001$, $\nu_{d} = \{1.0, 1.0, 0.2\}$, $m = 1$, $\alpha = 0.18$, $\pi_{W} = \pi_{W^{\text{in}}} = 0.1$ and $g(\cdot)$ set to a standard normal distribution. A main assumption of the ESN model is that $\bs{Y}_{t}$ follows a normal distribution as per equation \eqref{eqn:DESN1}. The logarithm of the data was used in order to achieve Gaussianity, thus it is assumed that the original data follows a log-normal distribution. In this Section, we also compare our ESN with Fixed Rank Kriging (FRK, \cite{cress08}), a popular space-time statistical model which uses basis decomposition and spatial random weights in order to reduce data dimensionality and account for nonstationarity (see supplementary material for a detailed technical explanation). We compare the forecasting skill of the DESN against a shallow ESN, ARFIMA, state-space model, HMM, FRK, LSTM and GRU in Section \ref{sec:AirPollForcs}, quantify the forecasting uncertainty and model the spatial dependence in Section \ref{sec:AirPollUncert}. Finally, in Section \ref{sec:AirPollExposure}, we interpolate forecasts, estimate the population exposure to excess levels of PM$_{2.5}$ and compare the PurpleAir results with the EPA station. The model diagnostics are detailed in the supplementary material.

\subsection{Forecasting Air Pollution}\label{sec:AirPollForcs}

The DESN, ARFIMA, and FRK forecasting methods for the same location from Figure \ref{fig:AirPollAdjustPI} can be seen in Figure S9. Similarly to the simulation study, these alternate models fail to produce accurate forecasts, hence underscoring 1) a nonlinear dynamics which cannot be captured by the linear structure of ARFIMA and 2) the lack of sufficient flexibility of the FRK approach for this application. Further evidence of the lack of fit can be assessed by computing the MSE and CRPS as in Section \ref{sec:sim}, with results in Table \ref{tbl:APMethodComp}. The ARFIMA model returns a median MSE value of 0.33 (IQR across sites 0.18), the state-space model a value of 0.49 (0.14). The HMM returns a MSE value of 0.37 (0.17), the FRK a MSE of 0.30 (0.30), and the LSTM and GRU return values of 0.38 (0.10) and 0.29 (0.10), respectively. The optimal DESN model generates a value of 0.14 (0.06), which exhibits the improvement in forecasting of the DESN against the other models. The DESN also outperforms the forecasts from shallow ESN, where this model returns a MSE of 0.19 (0.07). A similar ranking, with DESN outperforming the other models, can be seen with CRPS.
\begin{table}[tb!]
\begin{centering}
\begin{tabular}{||c|c|c|c||} 
\hline
Forecasting Method & MSE & CRPS \\ [1ex] 
\hline\hline
Deep ESN & 0.14 (0.06) & 0.21 (0.05) \\ 
\hline
Shallow ESN & 0.19 (0.07) & 0.25 (0.04) \\ 
\hline
ARFIMA & 0.33 (0.18) & 0.35 (0.09) \\ 
\hline
State-Space & 0.49 (0.14) & 0.44 (0.07) \\
\hline
Hidden Markov Model & 0.37 (0.17) & 0.36 (0.12) \\
\hline
Fixed Rank Kriging & 0.30 (0.30) & 0.32 (0.17) \\
\hline
Long Short-Term Memory & 0.38 (0.10) & 0.36 (0.05) \\
\hline
Gated Recurrent Unit & 0.29 (0.10) & 0.31 (0.05) \\[1ex]
\hline
\end{tabular}
\caption{Comparison of the forecasting methods in terms of median MSE and CRPS across all locations from the San Francisco air pollution data. In parenthesis the IQR across locations is reported.}
\label{tbl:APMethodComp}
\end{centering}
\end{table}

\subsection{Assessing the Forecast Uncertainty}\label{sec:AirPollUncert}

\begin{figure}[!tb]
\centering
\includegraphics[width = 15cm]{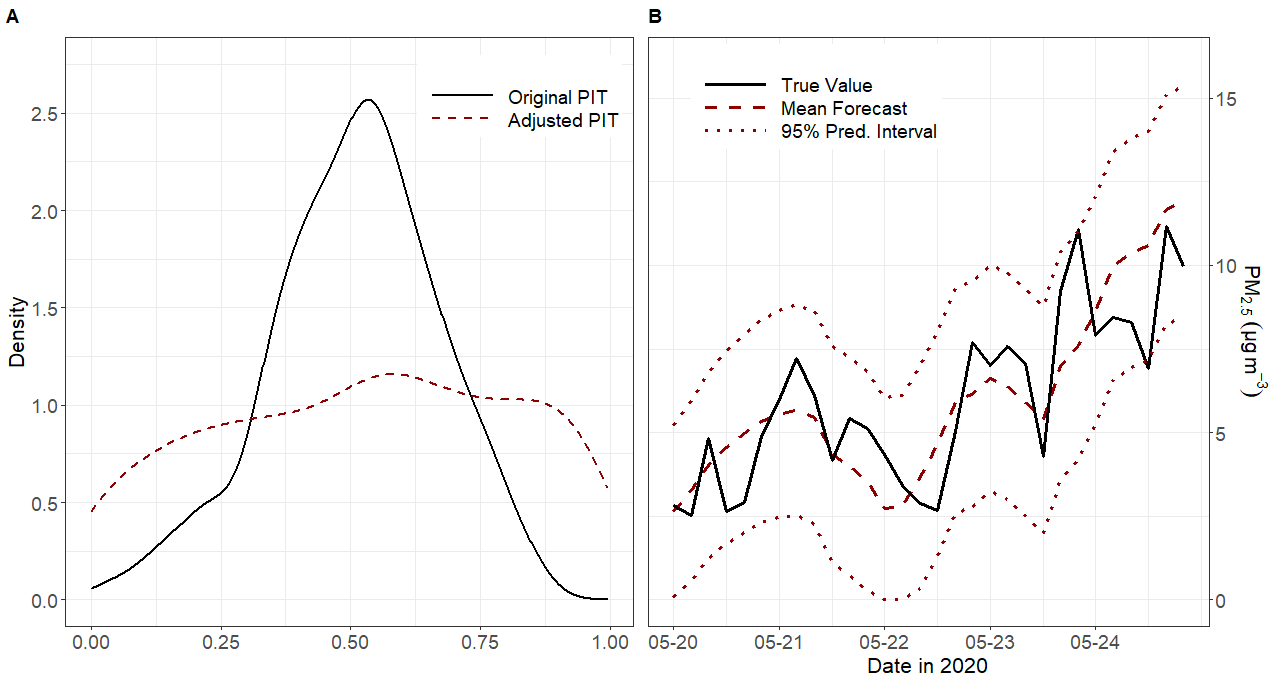}
\caption{For the sample location depicted in Figure \ref{fig:AirPollLocs}, (A) PIT for the uncalibrated and calibrated forecast and (B) long range calibrated forecasts with adjusted PIs.}
\label{fig:AirPollAdjustPI}
\end{figure}

Row 1 of Table \ref{tbl:APPICoverage} shows that, without any calibration of the forecasts, the original marginal 95\% PIs capture a median of 99.0\% of the data across all 44 locations with an IQR 0.5\%. After calibration, the median 95\% PIs coverage across all locations is 94.3\% (0.8\%), as shown in row 2, an improvement in coverage consistent with the results presented in the simulation study in Section \ref{sec:sim}. Additionally, Figure \ref{fig:AirPollAdjustPI}A shows how the adjustment method from Section \ref{sec:Uncertainty} calibrates the forecasts, as the PIT is closer to resembling a standard uniform distribution that the original uncalibrated forecast. Figure \ref{fig:AirPollAdjustPI}B shows the test data forecasts and adjusted 95\% prediction intervals for a sample location indicated in Figure \ref{fig:AirPollLocs}. From this plot, it can be seen that the post-adjustment PIs are able to capture the uncertainty in the data and become wider as we forecast further into the future, as we would expect. 

\begin{table}[tb!]
\begin{centering}
\begin{tabular}{||c|c|c|c|c||} 
\hline
Coverage & Method & 95\% & 80\% & 60\% \\ [1ex] 
\hline\hline
\multirow{2}{*}{Marginal} & Uncalibrated & 99.0 (0.5) & 96.7 (1.3) & 90.6 (2.1)\\
\cline{2-5}
& Calibrated & 94.3 (0.8) & 81.6 (2.3) & 65.4 (2.1)\\
\hline
\hline
\multirow{2}{*}{Difference} & Independent & 98.3 (0.8) & 93.1 (1.7) & 81.0 (3.9)\\
\cline{2-5}
& Dependent & 96.7 (1.7) & 88.3 (2.6) & 74.1 (5.2)\\
\hline
\hline
\multirow{2}{*}{Mean} & Independent & 44.0 & 28.1 & 16.5\\
\cline{2-5}
& Dependent & 94.4 & 80.0 & 64.6\\
\hline
\end{tabular}
\caption{Comparison of log PM$_{2.5}$ concentration forecast methods in terms of discrepancy between nominal and actual coverage. Rows 1 and 2 show the marginal coverage for uncalibrated and calibrated PIs. Rows 3 and 4 focus on the difference between forecasts in neighboring locations assuming independence and dependence (similarly to the simulation study in Section \ref{sec:sim}, only locations with absolute correlation between 0.4 and 0.6 were considered). Rows 5 and 6 compare the mean of the forecasts assuming independence and dependence. The median coverage is reported and in parenthesis the IQR is shown. 
}
\label{tbl:APPICoverage}
\end{centering}
\end{table}

In order to capture spatial dependence of the forecasts, we apply a convolution-based non-stationary model to the standardized residuals calculated from equation \eqref{eqn:StandResids}. This model allows the spatial structure to vary in space \citep{pacio06, risser17, Neto14}. More specifically, the correlation is generated as a result of the convolution at some fixed knots over the spatial domain. The correlation structure is formulated as in \cite{risser17}:
\begin{subequations}
\begin{flalign}
& {\hat{\textbf{C}}}_{\text{spatial}}(\bs{s},\bs{s}^{\prime}) = \frac{|\bs{V}(\bs{s})|^{1/4}|\bs{V}(\bs{s}^{\prime})|^{1/4}}{\left|\frac{\bs{V}(\bs{s})+\bs{V}(\bs{s}^\prime)}{2}\right|^{1/2}}k\left(\sqrt{Q(\bs{s},\bs{s}^{\prime})}\right),\nonumber\\
& Q(\bs{s},\bs{s}^\prime) = (\bs{s}-\bs{s}^\prime)^{\top}\left(\frac{\bs{V}(\bs{s})+\bs{V}(\bs{s}^\prime)}{2}\right)^{-1}(\bs{s}-\bs{s}^\prime),\nonumber
\end{flalign}
\end{subequations}

\noindent where $k(\cdot)$ is a correlation function, $\bs{V}(\cdot)$ is a matrix representing the local anisotropy and $Q(\bs{s},\bs{s}^\prime)$ is the Mahalanobis distance. In this work, $k(\cdot)$ is specified as the exponential function with a spatially-varying range and a fixed, unknown nugget. In order to specify $\bs{V}(\bs{s})$, we assume a mixture at some knots for $Z$ fixed locations $\{\textbf{b}_{z}; z=1,\dots,Z\}$. More specifically, we define
\begin{subequations}
\begin{flalign}
\begin{centering}
\bs{V}(\bs{s}) = \sum_{z=1}^{Z}w_{z}(\bs{s})\bs{V}_{z}, \qquad
w_{z}(\bs{s}) \propto \text{exp}\left\{-\frac{||\bs{s}-\textbf{b}_{z}||^{2}}{2\lambda_{z}}\right\},\nonumber
\end{centering}
\end{flalign}
\end{subequations}

\noindent where the weights, $w_{z}(\bs{s})$, are normalized. In this work $\lambda_{z}$ was fixed to be half the minimum distance between the knots. A similar approach is used to estimate the range parameter for the exponential correlation function. As discussed in \cite{risser17}, inference in performed through local likelihood.

In order to further improve our estimation of the spatial correlation structure, a new estimate is generated as a convex combination of the non-stationary correlation estimate, $\hat{\textbf{C}}_{\text{spatial}}$, and the empirical correlation, $\hat{\textbf{C}}$, a process known as generalized shrinkage \citep{fried07} which has been applied in previous works as a means to improve the estimated correlation structure \citep{Castruc18}. More specifically, the new correlation $\hat{\textbf{C}}^{*}(\delta_{c})$ is formulated as:
\begin{eqnarray}\label{eqn:ConvexCovar}
\hat{\textbf{C}}^{*}(\delta_{c}) = (1-\delta_{c})\hat{\textbf{C}}_{\text{spatial}} + \delta_{c}\hat{\textbf{C}},\nonumber
\end{eqnarray}

\noindent where $\delta_{c}\in(0,1)$ is a parameter to be estimated. In this work, $\delta_{c}$ was chosen such that the coverage of the grand mean of the data, in the log-scale, is as close to the specified PIs as possible. This correlation matrix $\hat{\textbf{C}}^{*}(\hat{\delta}_{c})$ therefore specifies the spatial dependence of the standardized residuals in equation \eqref{eqn:multinorm} and is used to calculate the PIs for the difference in neighboring locations and mean forecasts in Table \ref{tbl:APPICoverage} rows 3-6. Accounting for the spatial dependence among locations improves the coverage of the PIs for both the forecast difference of neighboring locations and forecast grand mean. Indeed, assuming independence among the locations results in an average over-coverage for the difference of 12.5\% versus an average over-coverage of 8.0\% when assuming dependence. When considering the coverage of the mean forecast, the spatial model dramatically improves the coverage. More specifically, the coverage discrepancy of the data, when assuming independence, is 48.8\% versus 1.7\% when assuming dependence. Even though the PIs are reported in the logarithmic scale, they are still interpretable in the original scale. In fact, the marginal coverage is unchanged since the logarithmic transformation is monotonic. The difference between neighboring elements and the mean in the logarithmic scale corresponds in the PM$_{2.5}$ scale to the ratio and the geometric mean, respectively. 

\subsection{Interpolation, Exposure and Comparison with EPA Data}\label{sec:AirPollExposure}

\begin{figure}[!tb]
\centering
\includegraphics[width = 12cm]{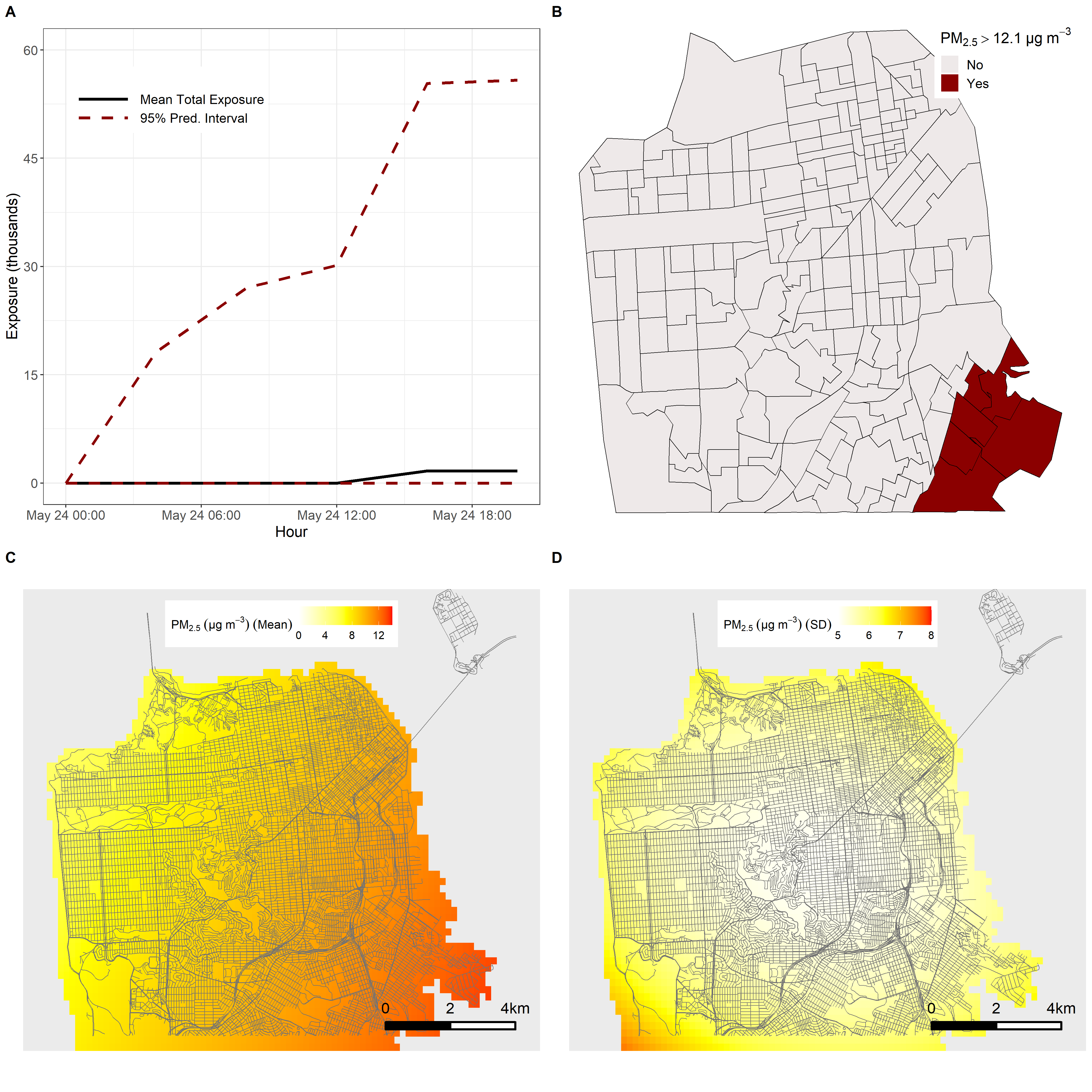}
\caption{(A) Time series of susceptible population exposed to levels of PM$_{2.5}$ concentration $> 12.1\upmu$g$\cdot m^{-3}$ across all San Francisco with PIs for 2020/05/24. (B) Population districts exposed to levels of PM$_{2.5}$ concentration $>12.1\upmu$g$\cdot m^{-3}$ at forecast time $T+30$, or 2020/05/24 at 20:00:00.  (C) Mean interpolated forecasts for all of San Francisco at time point $T+30$. (D) Standard deviation of interpolated forecasts for all of San Francisco at time point $T+30$.}
\label{fig:AirPollExposure}
\end{figure}

After the forecasts at the locations in Figure \ref{fig:AirPollLocs} were calibrated, maps of San Francisco were produced via interpolation using the spatial model in the previous section. The interpolated forecasts were then partitioned into 194 census tracts or districts, as determined by the United States Census Bureau, and the PM$_{2.5}$ value for each district was computed as the average of the interpolated forecasts within each district's borders. These forecasts were generated for all $\tau=30$ future 4-hour time points, which represents 2020/05/20 through 2020/05/24. Panel A in Figure \ref{fig:AirPollExposure} shows the total number of citizens exposed to `moderate' levels of air pollution (PM$_{2.5}>12.1 \upmu$g$\cdot m^{-3}$, \cite{epa}) for all forecasting points on 2020/05/24 with the corresponding PIs (these were the only time points with PM$_{2.5}>12.1 \upmu$g$\cdot m^{-3}$). We consider only susceptible citizens or those between the ages of 30 and 70 \citep{WHO14}, see supplementary material for the population data preprocessing. As seen in panel A, the number of individuals exposed increases before the early morning of 2020/05/24. Panel B in Figure \ref{fig:AirPollExposure} shows the population districts at forecast point $T+30$ (the time point with the maximum exposure) with PM$_{2.5}$ levels greater than 12.1 $\upmu$g$\cdot m^{-3}$. Here, the southeastern most districts of the city are exposed to `moderate' levels of air pollution, and this result coincides with the results presented in panel C where the interpolated forecasts are shown. Additionally, Panel D in Figure \ref{fig:AirPollExposure} shows the standard deviation of the interpolated forecasts. 

For comparison with EPA data, the DESN was implemented with the EPA monitoring station, with training and testing data on the same date range as the PurpleAir data. Since EPA data are collected at the daily level, the ESN was trained using only $T=140$ points and tested on the next $\tau=5$ points. The MSE of the EPA forecasts was 0.73, worse than the 0.14 with the PurpleAir data (row 1 of Table \ref{tbl:APMethodComp}). The inability of the DESN to produce as accurate forecasts with EPA data can be attributed to a smaller training set, but also to the lack of any spatial information. Indeed, for PurpleAir multiple locations are used to inform the forecasts, thereby borrowing spatial information. Additionally, none of the forecasts with the EPA data for all $\tau=5$ future points exceeded the threshold of $12.1 \upmu$g$\cdot m^{-3}$. Since this is the only air quality monitoring station within San Francisco, a map would trivially show the same value everywhere, thus losing information on local areas of high pollution as the ones presented in panels B and C of Figure \ref{fig:AirPollExposure}. 

\section{Conclusion}\label{sec:concl}

In this work we have introduced a new approach for forecasting air pollution in an urban environment and quantifying the associated uncertainty with a recurrent neural network. The large parametric space of NN-based models is not practical in applications with high-frequency data such as sub-daily air pollution. Therefore, we rely on an Echo-State Network (ESN), a faster alternative to standard dynamical NN models with sparse stochastic networks informed by a spike-and-slab prior and controlled by a low dimensional parameter space, thereby dramatically reducing the computational burden. The stochastic nature of the ESN allows for real time forecasts, but simultaneously introduces additional uncertainty that needs to be quantified. In this work, we propose a fast approach to calibrate the marginal uncertainty propagated by the stochastic network. Additionally, in order to produce multivariate (possibly spatial) forecasts, instead of embedding the dependence within the ESN in either the network or in the error structure (a task which would add computational complexity and hamper its practicality), we propose a \textit{post hoc} adjustment by modeling the dependence of the vector of marginally calibrated forecasts, either with penalized sparse correlation matrix estimation in a multivariate setting, or with a non-stationary spatial model in the case of urban air pollution. The proposed approach shows an appreciable improvement in the predictive abilities compared to standard forecasting strategies such as ARIMA, VAR, state-space models, fixed rank kriging, LSTM, and GRU and shows better-calibrated marginal and joint forecasts versus uncorrected or independent approaches. 

Another important contribution of this work is as a demonstration of the value of citizen science networks, and its potential to complement federal and non-profit efforts to map air pollution at larger scales, from national to global. This very recent source of data provides spatially-resolved information at urban scale, overcoming a major limitation of the sparse EPA network, satellite data and numerical simulations. This work represents one of the early efforts in assessing the added value of this new data in early warning systems on urban air quality, and some of our current work is expanding the methods presented here in assessing mortality from extreme events, such as wildfires \citep{she21}. While our work provides evidence of association of PurpleAir with EPA data, we are not in the position of validating our current estimate of exposure, as this task would require 1) access to local medical records; and 2) the attribution of any cardiorespiratory disease to a given event of air pollution rather than chronic exposure. 

While representing an opportunity, air pollution from citizen science data needs to be used with the awareness of potentially strong sampling bias due to income inequalities within the urban areas. While in less populous, relatively homogeneous cities, such as San Francisco, this may not be a substantial issue, considerably larger and more diverse cities would inevitably suffer from undersampling in areas of low income, where acquisition of such monitoring devices could not be as widespread. The sampling bias in citizen science projects due to underrepresentation of certain socio-economic groups is an emerging issue \citep{sor19}, and may be at least partially mitigated by federally funded, coordinated urban sensing projects such as the recent Array of Things \citep{cat17}.

\bibliographystyle{chicago}
\bibliography{references}

\end{document}